\documentstyle[prl,aps,multicol,psfig]{revtex}
\begin{document}
\title{Random field based model of mixed ferroelectrics phase diagram}
\author{M.D.Glinchuk, E.A.Eliseev}
\address{Institute for Problems of Materials Science, NASc of Ukraine,
Krjijanovskogo 3, 03180 Kiev, Ukraine}
\author{V.A.Stephanovich} 
\address{Institute of Semiconductor Physics NASc of Ukraine, Prospect Nauki 28,
Kiev, Ukraine and \\Institute of Mathematics University of Opole, Oleska 48, 45-052, Opole,
Poland}
\author{L.Jastrabik} 
\address{Institute of Physics, Academy of Sciences of the Czech Republic,
Na Slovance 2, Prague, Czech Republic}
\date{\today}
\maketitle

\begin{abstract}
The equations for phase transitions temperatures, order parameters and 
critical concentrations of components have been derived for mixed 
ferroelectrics. The electric dipoles randomly
distributed over the system were considered as a random field sources. We derive 
a random field distribution function for different orientations of the electric
dipoles with nonlinear and spatial correlation effects included.
The concentrational dependence of transition temperature for A$_{1-x}$B$_x$
mixed system has been calculated for A and
B being antiferroelectric and ferroelectric materials as well as for A
and B being ferroelectric and paraelectric. The numerical
calculations have been carried out for PbZr$_{1-x}$Ti$_x$O$_3$ and BaZr$_x$Ti$%
_{1-x}$O$_3$. The obtained results give a fairly good description of experimentally
observed phase diagrams of these mixed systems. We discuss the physical reasons of
strongly different behaviour of the systems related (in particular) to the
relaxor properties appearance in BaZr$_x$Ti$_{1-x}$O$_3$ at $x>0.27$ and
peculiar role of lead ions. We predict a transformation of any mixed 
system with ferroelectric (antiferroelectric) and paraelectric component 
into relaxor material at some
large enough concentration of paraelectric component.
\end{abstract}

\begin{multicols}{2}

\narrowtext

\section{Introduction}

There has been considerable amount of attention of scientists to mixed systems 
of different types like ferroelectric and
antiferroelectric, ferroelectric and paraelectric etc due to their unusual 
physical properties, utilizable in many applications. A classic
example of such a system is PbZr$_{1-x}$Ti$_x$O$_3$ (PZT) with broad range
of its technical usage \cite{bt}, including novel branches
of electronics \cite{gu}. Most of the important peculiarities occur in
the special regions of concentrations in the vicinity of
different phase boundaries on phase diagram. In particular, for PZT the most
interesting range is the morphotropic one with coexistence of ferroelectric phases with
two different symmetries and antiferroelectric-ferroelectric phase
boundary \cite{gu,ha}. The characteristic features of phase
diagram and properties anomalies depend essentially on the type
of solid solution components (ferroelectric, antiferroelectric or
paraelectric). It is especially interesting from this point of view to
compare phase diagrams of PZT (solid solution of antiferroelectric PbZrO$_3$
and ferroelectric PbTiO$_3$) and BaZr$_x$Ti$_{1-x}$O$_3$ (BZT, which is
solid solution of ferroelectric BaTiO$_3$ and paraelectric BaZrO$_3$ \cite{ke}. 
The latter clearly demonstrates a peculiar role of lead ions in phase
transitions of perovskite structure materials (see e.g. \cite{lg}). 
Due to different states (e.g. ferroelectric, paraelectric etc) of PZT and BZT 
components, their phase diagrams are strongly different (see e.g. \cite{cr1,far1}).
Namely, several phases with long range order exist in PZT while BZT exhibits
relaxor behaviour beginning at $x=0.27$. To find out the physical mechanisms "responsible"
for actual phase diagram formation in mixed system, the theoretical
calculations seem to be extremely desirable. Recently, the random field model for calculation
of physical properties of mixed system had been proposed \cite{mu3}. 
In the present work this
model is expanded by taking into account the nonlinear and correlation effects 
as well as different orientations of the dipoles. This made it possible to
calculate the concentration dependence of transition temperature, determine the symmetry
of order parameters (and obtain the change of mixed system symmetry). This model also
permits to obtain the range of parameters for existence of morphotropic region 
and appearance of glassy state. We apply this
essentially improved model to calculations of phase diagrams of PZT and BZT. 
The theory describes adequately the observed phase diagram in
these materials.

\section{Order parameters}

In the case of solid solutions of antiferroelectric and ferroelectric
materials we have to consider three order parameters. Namely, in two-
sublattice model for antiferroelectric component there are ferroelectric $%
L_{2F}$ and antiferroelectric $L_{2A}$ order parameters that describe, respectively,
the homogeneous and inhomogeneous displacements of ions. The third
order parameter $L_{1F}$ is related to another component of solid solution.
Keeping in mind the key role of electric dipoles in the system and their
ability to be oriented by electric field $E$ we write the order parameters
in the form inherent for  Ising model\cite{vaks}:
\begin{mathletters}
\begin{eqnarray}
&&L_{1F}^{mf} = \tanh\left( (d_1^{*}E+T_{1F}L_{1F})/T\right)   \label{1a} \\
&&L_{2F}^{mf} =\frac 1Z [\sinh\left( 2(d_2^{*}E+T_{2F}L_{2F})/T\right) ]  \label{1b} \\
&&L_{2A}^{mf} =\frac 1Z [\sinh\left( 2T_{2A}L_{2A}/T\right) ] \label{1c}\\
&&Z=\cosh \left(2(d_2^{*}E+T_{2F}L_{2F})/T \right) +\cosh(2T_{2A}L_{2A}/T) \nonumber 
\end{eqnarray}
\end{mathletters}
Here $d_1^{*}$ and $d_2^{*}$ are effective dipole moments related to the
first ($x=1$) and the second ($x=0$) components; $T_{1F}$ is ferroelectric
phase transition temperature of the first component and $T_{2F}$, $T_{2A}$
are the transition temperatures of ferroelectric and antiferroelectric phases
of the second component of solid solution. Superscript ''mf'' in $L$ shows
that Eqs.(1) are written in a mean field approximation.

In the mixed system the positions of the dipoles of both components become
random so that they can be considered as a random field sources. In this case
order parameters given by Eqs. (1) have to be averaged with the random
field distribution function $f(E,L_{1F},L_{2F})$. Note, that in general case
electric field $E$ is the sum of external and internal field, but hereafter
we suppose the absence of external field. Allowing for nonlinear and spatial
correlation effects contribution into the distribution function, \cite{far3}
order parameters for the mixed system can be written as:
\begin{mathletters}
\begin{eqnarray}
&&L_{1F} =\int \tanh(d_1^{*}\varphi _1({\bf Ee}_1)/T)f({\bf E}%
,L_{1F},L_{2F})d^3E,  \label{2a} \\
&&L_{2F} =\int \frac{\sinh(d_2^{*}\varphi _2({\bf Ee}_2)/T)f({\bf E}%
,L_{1F},L_{2F})d^3E}{\cosh(d_2^{*}\varphi _2({\bf Ee}_2)/T)+\cosh(2T_{2A}L_{2A}/T)}%
,  \label{2b} \\
&&L_{2A} =\int \frac{\sinh(2T_{2A}L_{2A}/T)f({\bf E},L_{1F},L_{2F})d^3E}{%
\cosh(d_2^{*}\varphi _2({\bf Ee}_2)/T)+\cosh(2T_{2A}L_{2A}/T)},  \label{2c} \\
&&\varphi _i(E) =E(1+\alpha _3^{(i)}E^2),\quad {\bf e}_i=\frac{{\bf d}_i^{*}%
}{\left| {\bf d}_i^{*}\right| }.  \label{2d}
\end{eqnarray}
\end{mathletters}
Here $\varphi _i(E)$ is the nonlinear function of the field, which has the form of 
infinite series in odd powers of $E$ in the lattices with a center of inversion 
in paraelectric phase. We keep in (\ref{2d}) only first nonvanishing nonlinear term with
the coefficient $\alpha _3$. Also, $f(E,L_{1F},L_{2F})$ is a linear random field
distribution function. It is seen, that in the case when both components are
ferroelectrics, i.e. $L_{2A}=0$, Eq.(\ref{2b}) has the same form as Eq.(\ref{2a}) as 
it has to be expected, but with different parameters. Therefore Eqs.(2)
describe a quite general case. They give the order parameters dependence on
the characteristics of mixed system components and their molar fractions via
the random field distribution function. We would like to emphasize that the
parameters $L$ are the fraction ($0\leq L\leq 1$) of coherently oriented
dipoles of the mixed system components. Polarization ${\bf P}$ as the actual
order parameter of mixed system can be expressed via $L_i$ as:

\begin{equation}
{\bf P}=\frac x{a_1^3}L_{1F}{\bf d}_1^{*}+\frac{1-x}{a_2^3}L_{2F}{\bf d}%
_2^{*},  \label{3}
\end{equation}
where $a_1$, $a_2$ and $x$, $1-x$ are respectively lattice constants and
molar fractions of the first and the second components in the solid solution
with chemical formula A$_{1-x}$B$_x$. Note, that $L_{2A}$ contributes neither to
mixed system polarization nor to the random field distribution function as
it has to be expected. The orientation of mixed system polarization is
related to the vector sum of the dipoles in accordance with Eq.(\ref{3}). To derive
this sum coefficients, one has to calculate $L_{1,2F}$ on the base of Eqs.(2)
which depend on the form of function $f(E,L_{1F},L_{2F})$.

\section{The random field distribution function}

As we have seen above, to calculate the nonlinear distribution function of random fields,
it is sufficient to calculate the linear distribution
function (i.e. that without nonlinear and spatial correlation effects). For latter purpose
all kind of dipoles can be considered as independent sources of random field. Therefore, the
distribution function of the mixed system is a convolution of the two types
of dipoles distribution functions \cite{kn}. Gaussian approximation for these functions 
leads to the following expression for mixed system distribution function:
\end{multicols}
\widetext
\noindent\rule{20.5pc}{0.1mm}\rule{0.1mm}{1.5mm}\hfill
\begin{equation}
f({\bf E})=\frac 1{(2\pi )^3}\int \exp \left( i{\bf \rho }({\bf E}-{\bf E}%
_0)-x\Delta E_1^2({\bf e}_1{\bf \rho })^2-(1-x)\Delta E_2^2({\bf e}_2{\bf %
\rho })^2\right) d^3\rho .   \label{4}
\end{equation}
\hfill\rule[-1.5mm]{0.1mm}{1.5mm}\rule{20.5pc}{0.1mm}
\begin{multicols}{2}
\narrowtext
\noindent
\begin{equation}
{\bf E}_0=x\frac{T_{1F}}{d_1^{*}}L_{1F}{\bf e}_1+(1-x)\frac{T_{2F}}{d_2^{*}}%
L_{2F}{\bf e}_2 .  \label{5}
\end{equation}

Here ${\bf E}_0$ is a mean field, $\Delta E_1$ and $\Delta E_2$ are the
halfwidths of the distribution functions induced by $d_1^{*}$ and $d_2^{*}$
dipoles respectively. They can be written in the form \cite{mu1}:

\begin{equation}
\Delta E_1^2=\frac{16\pi }{15}\frac{d_1^{*2}}{\varepsilon _1^2r_{c1}^3a_1^3}%
,\quad \Delta E_2^2=\frac{16\pi }{15}\frac{d_2^{*2}}{\varepsilon
_2^2r_{c2}^3a_2^3},  \label{6}
\end{equation}
where $\varepsilon _1$ and $r_{c1}$, $\varepsilon _2$ and $r_{c2}$ are
dielectric permittivities and correlation radii of the solid solution
components.

One can see from Eqs. (\ref{4}), (\ref{5}), (\ref{6}) that mixed system 
distribution function
depends on the components concentrations, order parameters $L_{1F}$ $L_{2F}$%
, transition temperatures $T_{1F}$ $T_{2F}$, their dipole moments and other
physical parameters.

\section{Phase diagram. General equations.}

The phase diagram has to reflect the variation of the transition
temperatures of different phases with the increasing of the fraction of the
components as well as the phases symmetry changing. All this information can
be obtained by solving the equations (2) with respect to Eqs. (\ref{4}),
(\ref{5}). Substitution of Eq.(\ref{4}) into (2) leads to six-fold integrals. 
They can
be simplified due to the dependence of all the integrands on the scalar
product $({\bf Ee}_i)$, i.e. only on the parallel to ${\bf e}_i$ field
component. This permits to integrate out two other field components and to
perform the integration over $d^3\rho $. This finally yields:
\end{multicols}
\widetext
\noindent\rule{20.5pc}{0.1mm}\rule{0.1mm}{1.5mm}\hfill
\begin{eqnarray}
L_{1F} &=&\int\limits_{-\infty }^{+\infty }\tanh\left( d_1^{*}\varphi
_1(E)/T\right) \exp \left( -\left( \frac{E-E_{01}}{2\Delta _1}\right)
^2\right) \frac{dE}{2\sqrt{\pi }\Delta _1},  \nonumber \\
L_{2F} &=&\int\limits_{-\infty }^{+\infty }\frac{\sinh\left( 2d_2^{*}\varphi
_2(E)/T\right) }{\cosh\left( 2d_2^{*}\varphi _2(E)/T\right) +\cosh\left(
2T_{2A}L_{2A}/T\right) }\exp \left( -\left( \frac{E-E_{02}}{2\Delta _2}%
\right) ^2\right) \frac{dE}{2\sqrt{\pi }\Delta _2},  \label{7} \\
L_{2A} &=&\int\limits_{-\infty }^{+\infty }\frac{\sinh\left(
2T_{2A}L_{2A}/T\right) }{\cosh\left( 2d_2^{*}\varphi _2(E)/T\right) +\cosh\left(
2T_{2A}L_{2A}/T\right) }\exp \left( -\left( \frac{E-E_{02}}{2\Delta _2}%
\right) ^2\right) \frac{dE}{2\sqrt{\pi }\Delta _2},  \nonumber
\end{eqnarray}
\hfill\rule[-1.5mm]{0.1mm}{1.5mm}\rule{20.5pc}{0.1mm}
\begin{multicols}{2}
\narrowtext
\noindent
where $E_{0i}=({\bf E}_0{\bf e}_i)$ ($i=1,2$), so that $E_{0i}$ depends on
the angle $\theta $ between the directions of the two types of dipoles,
because $({\bf e}_1{\bf e}_2)=\cos ({\bf e}_1,{\bf e}_2)\equiv \cos \theta $%
. The parameters $\Delta _i$ also depend on this angle, namely:

\begin{eqnarray}
\Delta _1 &=&x(\Delta E_1)^2+(1-x)(\Delta E_2\cos \theta )^2,  \label{8} \\
\Delta _2 &=&x(\Delta E_1\cos \theta )^2+(1-x)(\Delta E_2)^2.  \nonumber
\end{eqnarray}

The Eqs.(\ref{7}) are the final form of the equations for the
order parameters $L_{2A}$, $L_{2F}$, $L_{1F}$ dependence on the molar
fraction $x$ and via them polarization of the system (see Eq.(\ref{3})). The
transition temperature $T_C$ to the ferroelectric phase and the transition
temperature $T_A$ to antiferroelectric one in the mixed system can be
derived from Eq.(7) in the limit of zeroth order parameters. One
can obtain from Eq.(7) the following system of equations:
\begin{mathletters}
\begin{eqnarray}
L_{1F} &=&\frac{T_{1F}}{T_C}\left( xL_{1F}+(1-x)\frac{\cos \theta }{p\lambda 
}L_{2F}\right) I_1(T_C),  \label{9a} \\
L_{2F} &=&\frac{T_{2F}}{T_C}\left( p\lambda \cos (\theta
)xL_{1F}+(1-x)L_{2F}\right) I_2(T_C),  \label{9b} \\
L_{2A} &=&\frac{T_{2A}}{T_A}(1-x)L_{2A}I_2(T_A),  \label{9c}
\end{eqnarray}
\end{mathletters}
where

\begin{equation}
I_i(T)=\frac 1{\sqrt{\pi }}\int\limits_{-\infty }^{+\infty }\frac{\left(
1+3\alpha _i(Q_iu)^2\right) \exp \left( -\left( u/2\right) ^2\right) }{%
ch\left( Q_iu\left( 1+\alpha _i(Q_iu)^2\right) T_{iF}/T\right) }du.  \label{10}
\end{equation}

Here dimensionless variables are introduced:
\end{multicols}
\widetext
\noindent\rule{20.5pc}{0.1mm}\rule{0.1mm}{1.5mm}\hfill
\begin{eqnarray}
p &=&\frac{d_2^{*}}{d_1^{*}},\quad \lambda _F=\frac{T_{1F}}{T_{2F}},\quad
\alpha _i=\alpha _3^{(i)}\left( \frac{T_{iF}}{d_i^{*}}\right) ^2,\quad q_i=%
\frac{d_i^{*}\Delta E_i}{T_{iF}},\quad \lambda _A=\frac{T_{2A}}{T_{2F}}, 
\label{11} \\
Q_1^2 &=&xq_1^2+(1-x)\left( \frac{q_2\cos \theta }{p\lambda _F}\right)
^2,\quad Q_2^2=x\left( q_1p\lambda _F\cos \theta \right) ^2+(1-x)q_2^2. 
\nonumber
\end{eqnarray}

One can see that the temperature $T_A$ follows from Eq.(9c), while the
temperature $T_C$ can be derived from Eqs.(9a), (9b). The solution of these
equations has the form:
\begin{mathletters}
\begin{eqnarray}
\tau _C &\equiv &\frac{T_C}{T_{2F}}=\frac 12\left( x\lambda _FI_1(\tau
_C)+(1-x)I_2(\tau _C)\pm \right.   \label{12a} \\
&&\left. \pm \sqrt{\left( x\lambda _FI_1(\tau _C)\right) ^2+\left(
(1-x)I_2(\tau _C)\right) ^2+2\cos (2\theta )\lambda _Fx(1-x)I_1(\tau
_C)I_2(\tau _C)}\right) ,  \nonumber
\end{eqnarray}

\begin{equation}
\tau _A\equiv \frac{T_A}{T_{2A}}=(1-x)I_2(\tau _A).  \label{12b}
\end{equation}
\end{mathletters}
\hfill\rule[-1.5mm]{0.1mm}{1.5mm}\rule{20.5pc}{0.1mm}
\begin{multicols}{2}
\narrowtext
\noindent
The solution of Eqs (12) gives the dependences of $T_C$
and $T_A$ on molar fractions and material parameters of the mixed system
components (see (11)).

\section{Comparison of the theory and experiment}

\subsection{Phase diagram of PbZr$_{1-x}$Ti$_x$O$_3$}

This solid solution components are antiferroelectric PbZrO$_3$ with
transition temperature $T_{2A}=503$ K (the value $T_{2F}$ $\approx T_{2A}$
\cite{vaks}) and ferroelectric PbTiO$_3$ with transition temperature from
paraelectric phase to tetragonal ferroelectric phase $T_{1F}=763$ K.
Both components have electric dipoles randomly distributed in the mixed
system. In accordance with the components symmetry one can suppose that $%
d_1^{*}\parallel [001]$ and $d_2^{*}\parallel [111]$ types of directions.
Therefore parameters $\lambda _F\approx $ 1.516, $\lambda _A=1$ and $\cos
\theta =1/\sqrt{3}$. Other parameters were obtained from the fitting with
observed phase diagram of PZT. We begin with the fitting of $T_A$ from
Eq.(12b). The numerical calculations of the integral $I_2$ were performed at 
$\alpha _2=0.3$, $q_2=2.9$. The dependence of $T_A$ on molar fraction $x$ is
depicted by dashed line in Figure 1. 
\begin{figure}
\vspace*{-4mm}
\centerline{\centerline{\psfig{figure=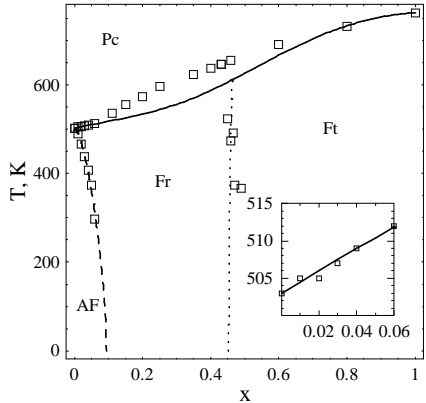,width=0.8\columnwidth}}}
\vspace*{-3mm}
\caption{ Phase diagram of PbZr$_{1-x}$Ti$_x$O$_3$. Open squares are the
experimental data \protect\cite{ja}. Calculated transition temperatures
are the following: solid line represents the transition from paraelectric
phase (Pc) to ferroelectric phases (Fr, Ft), dashed line represents the
transition from antiferroelectric phase (AF) to ferroelectric rhombohedral
phase (Fr), dotted line represents the transition from (Fr) phase to
ferroelectric tetragonal phase (Ft). The transition between Pc and Fr phases
at small fractions $x$ is depicted in inset.}
\end{figure}
One can see that $T_A$ decreases with $x
$ increase and $T_A=0$ K at $x=x_C=0.093$ (where $x_C$ is a critical
fraction at which antiferroelectric phase disappears). It is seen that the
theory gives reasonable fit to experimental data shown by open squares in
Figure 1. The transition temperature $T_C$ calculated on the
base of Eq.(12a) increases with $x$ increase (see solid line in Figure 1).
Our theory is undoubtedly valid at small $x$ (see inset to Figure 1).
The fitting of $T_C$ for all the range of $x$ were performed by varying of $p
$, $q_1$, $q_2$, $\alpha _1$, $\alpha _2$. The best fit was achieved at $%
p=0.828$, $q_1=0.239$, $q_2=0.364$, $\alpha _1$ = 3.9, $\alpha _2$ = 4.3. As
it follows form Figure 1, the fitting is also good for $x>0.6$, i.e. in the
region enriched by titanium. For intermediate molar fractions $0.1<x<0.6$
the accuracy of the fitting is not so good as it is in the other regions. 
\begin{figure}
\vspace*{-4mm}
\centerline{\centerline{\psfig{figure=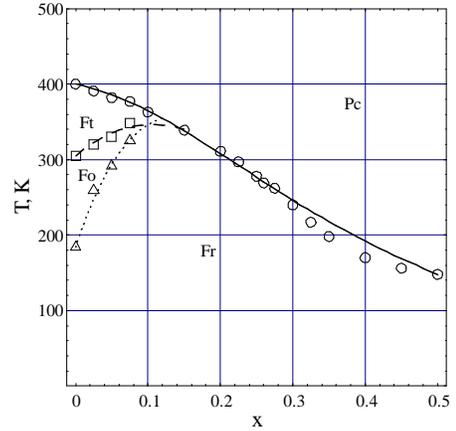,width=0.8\columnwidth}}}
\vspace*{-3mm}
\caption{ Phase diagram BaZr$_x$Ti$_{1-x}$O$_3$. Open circles, squares and
triangles are the experimental data from \protect\cite{far1}. Calculated
transition temperatures are the following: dashed line represents the
transition between tetragonal (Ft) and orthorhombic (Fo) ferroelectric
phases, dotted line represents the transition from (Fo) phase to
ferroelectric rhombohedral phase (Fr), solid line represents the transition
from paraelectric cubic phase (Pc) to (Ft, Fr) phases.}
\end{figure}
To our mind this deviation is related to the existence (in this interval of $x$)of
additional order parameters originated from an improper ferroelectric phase
transition. This transition occurs between high and low temperature rhombohedral phases.
In the high temperature phase the spontaneous tilting of 
the oxygen octahedra causes the phase
transition and contributes to the spontaneous polarization \cite{cr1}.
The symmetry of different ferroelectric phases and morphotropic
region with both symmetries coexistence (see dotted line in Figure 1)
was calculated on the base of Eqs. (3), (7), (8) with the same set of
parameters as that used for the transition temperatures fitting. It is clear from
Figure 1 that the morphotropic region lies between $x=0.453$ at $T=0$ K and $%
x=0.463$ at $T=611$ K. Therefore the calculations describe adequately the
observed phase diagram \cite{ja} represented by open
squares in the Figure 1. Note that the best fit was obtained in the
assumption that the halfwidths $\Delta _i$ of the distribution functions
for each component are determined by the other component of solid solution.
This confirms the supposition that the sources of the random field of the
first component destroy the long- range order of the second component and
vise versa.

\subsection{Phase diagram of BaZr$_x$Ti$_{1-x}$O$_3$}

The main component of this solid solution is the ferroelectric material BaTiO%
$_3$ which is known to have three phase transitions at $T_{2F}^{(1)}=400$ K, 
$T_{2F}^{(2)}=305$ K, $T_{2F}^{(3)}=184$ K, to ferroelectric phases with
tetragonal, orthorhombic and rhombohedral symmetries respectively. BaZrO$_3$
is paraelectric at all temperatures \cite{ke}. Phase diagram of BZT
differs strongly from that of PZT. To describe it we proposed the
following model.

We assume that in mixed BZT system zirconium ions can be shifted so that
they supply the random electric dipoles to BaTiO$_3$ component. These
dipoles are the main sources of the random field. This field distribution
function halfwidth has to be larger than the mean field produced by
zirconium dipoles because BaZrO$_3$ is paraelectric. Taking all these
arguments into consideration we describe the phase diagram of BZT system
with the help of Eqs.(3), (7), (8) with the following set of parameters:

\begin{eqnarray}
T_{1F} &=&250~K,\quad q_1=0.6,\quad q_2=0,\quad \alpha _1=0;  \nonumber \\
\alpha _2^{(1)} &=&0.3,\quad p^{(1)}=2;\quad \alpha _2^{(2)}=2,\quad
p^{(2)}=1.63;  \label{13} \\
\alpha _2^{(3)} &=&8,\quad p^{(3)}=1.25.  \nonumber
\end{eqnarray}

Note, that is would be more accurate to write $T_{1mf}$ instead of $T_{1F}$
because there is no actual ferroelectric phase transition in BaZrO$_3$.

Allowing for known symmetry of three BaTiO$_3$ ferroelectric phases, one can
obtain $\cos (\theta ^{(1)})=1$, $\cos (\theta ^{(1)})=1/\sqrt{2}$, $\cos
(\theta ^{(3)})=1/\sqrt{3}$. The results of the calculation are shown in
Figure 2. One can see that the theory describes the observed
phase diagram (see \cite{far1} and references therein) quite well. Note, that
the accuracy of fitting of experimental points by dashed and dotted lines at 
$x>0.12$ is about 10\%.

In Figure 3 we represented BZT order parameters calculated with the
parameters (13) at $T=0$ K. It is easy to check that in this limit $L_1=L_2$%
. One can see that at $x>0.3$ the fraction of coherently oriented dipoles $L$
is less than 0.9 which corresponds to mixed ferroglass phase \cite{far2},
 the critical fraction for the dipole glass appearance ($L=0$)
being about $x_c\approx 0.82$. The relaxor behaviour (e.g. Vogel-Fulcher law
in dynamic permittivity) was observed recently at $x\geq 0.27$ (see \cite{far1}
and references therein).

The value of $x_c$ for the transition to dipole glass state is a prediction of
the theory. Unfortunately, the available experimental data cover the range up
to $x=0.5$. The clarification of dipole glass state existence may be performed
in solid solution of BaTiO$_3$ and BaZrO$_3$ only under the condition of the
components solubility in all the concentration range.
\begin{figure}
\vspace*{-4mm}
\centerline{\centerline{\psfig{figure=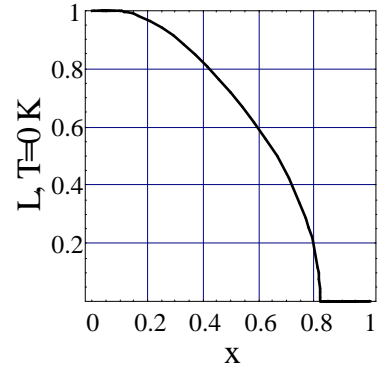,width=0.8\columnwidth}}}
\vspace*{-3mm}
\caption{ The dependence of BaZr$_x$Ti$_{1-x}$O$_3$ order parameters at $T=0$
K versus the molar fraction $x$.}
\end{figure}

\section{Discussion}

In the proposed model two components of the mixed system were considered as
the materials with electric dipoles ${\bf d}_1$ and ${\bf d}_2$. The dipoles
tend to order the system along their directions so that the competition
between different ordering directions is the main mechanism defining 
phase diagram of mixed system.
Without this competition (when there is only one type of dipoles) it can
be ferroelectric (PbTiO$_3$, BaTiO$_3$),
antiferroelectric (PbZrO$_3$) or paraelectric (BaZrO$_3$) phases. The
difference between properties of the materials with Pb or Ba ions is related
to peculiar role of lead ions in the phase transitions. In perovskite
structure ABO$_3$ all A ions, but Pb, give almost no contribution to a
lattice polarization. In contrast to this Pb ions contribution (e.g. in PbTiO$%
_3$) is the main one (see e.g. \cite{lg} and ref. therein).
Opposite displacements of lead ions in PbZrO$_3$ are known to be the
characteristic feature of antiferroelectric phase in PbZrO$_3$. In our model 
${\bf d}_i$ represents the resultant dipole moment of a lattice unit cell.
It is a vector sum of Pb and Ti ions displacements in the case of
PbTiO$_3$ or Ti ions in BaTiO$_3$, all the displacements being considered
relatively oxygen cage. Therefore the extraction of lead contribution can be
made only on the base of independent microscopic calculations or
measurements of $d_1^{*}/d_2^{*}$ ratio which is fitting parameter in our
model. The estimation of this ratio with the help of known displacements
(see e.g. \cite{lg}) of Ti ions in PbTiO$_3$ and Zr ions in
PbZrO$_3$ gives $p=d_2^{*}/d_1^{*}\approx 0.8$ while our fit gives $p=0.828$
(see section 5.1). Since nothing is known about Zr displacements in BaZrO$_3$
it is resonable to calculate, e.g., the ratios of $k_1=p_1/p_3$ and $%
k_2=p_2/p_3$, which are independent on dipole moment of Zr ions. Under
supposition that $k_1$ and $k_2$ can be estimated via the ratio of
polarizations in ferroelectric phases of BaTiO$_3$ (see
e.g. \cite{iosh}), one obtains $k_1=1.8$, $k_2=1.4$ while in our
model $k_1=1.6$, $k_2=1.3$ (see (13)).  The values of $p$ in
(13) show that Zr displacement in BaZrO$_3$ is two times smaller, than that of Ti
in tetragonal phase of BaTiO$_3$. Therefore our model leads to reasonable
values of the ions dipole moments both in PZT and BZT. It is actually possible 
to estimate fitting parameters with the help of Eqs.(6) provided that correlation radii and
dipole moments are known. The parameters related to nonlinearity
coefficients play an important role in description of peculiar form of BZT
phase diagram because they are "responsible" for $T_c(x)$ maxima. Although the
relation between $\alpha _2^{(i)}$ is qualitatively the same as that between
nonlinearity constants $\alpha $ in different phases of BaTiO$_3$ (namely $%
\alpha _2^{(1)}<\alpha _2^{(2)}<\alpha _2^{(3)}$ (see (13)) and $\alpha
^{(1)}<\alpha ^{(2)}<\alpha ^{(3)}$ \cite{iosh}), quantitative
estimation of $\alpha _2^{(i)}$ seems to be complex problem and the
independent measurements of these parameters are desirable. Therefore these
parameters are the actual fitting ones while other parameters could be
estimated {\em a priori}. Another interesting feature of BZT phase diagram is
appearance of relaxor behaviour at $x\geq 0.27$ while in PZT nothing of this
kind is known. This may be related to the fact that in PbTiO$_3$ (due to lead
ions displacements) spontaneous polarization is more than 2 times larger than
that in BaTiO$_3$ \cite{lg}. As a result the random field
induced by Zr ions appeared unable to destroy strong ferroelectric order in
PbTiO$_3$ whereas in BaTiO$_3$ this field can destroy weaker
ferroelectric long range order at large enough Zr ions concentration.

More generally we can assert that for a mixed system containing
ferroelectric and paraelectric components, it has to be the concentration
range where the system transforms into relaxor. This statement follows
from the fact that paraelectric component contribution to mean field $E_0$
is rather small while it completely defines the distribution function
half-width ($q_2=0$, see (13)). The contribution of the ferroelectric
component to $E_0$ decreases (see Eq. (5) with subscripts 1 and 2 corresponding
to paraelectric and ferroelectric component respectively and $T_{2A}=0$, $L_{2A}=0$ in Eqs.
(7)). So, the increase of paraelectric component concentration leads to mean field
decrease and to increase of distribution function half-width $\Delta $ (see
Eqs. (6), (8)). This must result into $E_0/\Delta (x)$ decrease.
In supposition that the state of a system (paraelectric (PE),
ferroelectric (FE), dipole glass (DG), mixed ferroglass (FG) where
FE long range order coexists with DG short range order) 
strongly depends on ratio $E_0/\Delta $ (see Figure 4), one can
conclude that as $x$ increases, the system passes from FE to FG and than to
DG state at some low temperature region (see arrows in Figure 4). Both FG
and DG states are known to be characteristic feature of relaxor materials
\cite{far2}. The relaxor systems exhibit 
nonergodic behaviour just in these phases. 
Vogel-Fulcher law which describes temperature
dependence of dynamic dielectric susceptibility of relaxors can be related
to the distribution of random electric fields in the mixed system \cite{mu1}. 
The calculations of concentrational dependence of (BaTiO$_3$)$_{1-x}$(BaZrO$_3$)$_x$
order parameter made it possible to obtain the
concentration ($x\approx 0.3$, see Figure 3) at which mixed system
transforms into relaxor. It follows from Figure 3 that BZT is in FG phase at 
$0.3<x<0.8$ and could be in DG state at $x\geq 0.8$, the latter being dependent on
existence of this $x$ region in BZT mixed system. The existence of DG state
and relaxor properties in another mixed system (BaTiO$_3$)$_{1-x}$(SrTiO$_3$)%
$_x$ (BST) at $x\geq 0.9$ \cite{lem}
confirms the generality of the statement about transformation of mixed
ferroelectric-paraelectric system into relaxor for some concentration
range. This transformation seems to exist also in the mixed system consisting of
antiferroelectric (component 2) and paraelectric (component 1).
Really, from Eqs. (5), one can expect the decrease of $E_0/\Delta $ so 
the system at $x$ increase passes from FE to FG and DG states (see 
the arrow in Figure 4). However, contrary to the case of ferroelectric-paraelectric
mixed system, where $T_{2F}$ is actual temperature of ferroelectric
phase transition, in the above considered case $T_{2F}$ is characteristic of
"imaginary" ferroelectric phase following from two sublattice model of
antiferroelectrics.  Its value is close (although little lower)
to $T_{2A}$ \cite{vaks}.
\begin{figure}
\vspace*{-4mm}
\centerline{\centerline{\psfig{figure=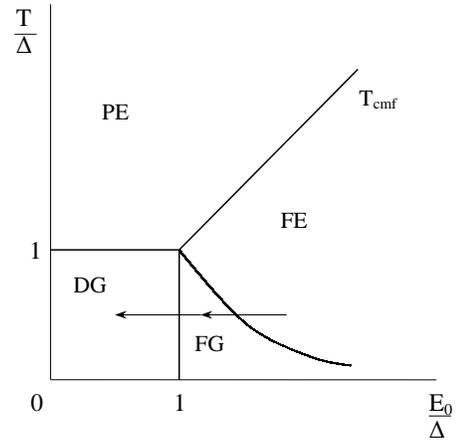,width=0.8\columnwidth}}}
\vspace*{-3mm}
\caption{Scheme of the disordered system phase diagram \protect\cite{by}}
\end{figure}

\section{Conclusion}

We propose the random field based theory for calculation 
of phase diagram of mixed ferroelectrics and apply it to PZT and BZT materials. It has
been shown that proposed theoretical approach describes both qualitatively
and quantitatively the observed phase diagrams, including relaxor behaviour
in BZT at $x>0.3$ (the measured fraction $x\approx 0.27$).
This discrepancy may be related to the model assumption that electric
dipoles of titanium and zirconium ions are the main sources of random
fields. Zirconium ions can be considered as dilatation centers or
elastic dipoles which are known to destroy a long-range order
leading to relaxor properties appearance \cite{far2}. The
calculations of latter property and of the 
contribution of oxygen octahedra tilting to the
polarization of PZT are in progress now.

\acknowledgments

Authors are grateful to Prof. T.Egami for fruitful discussions of the results.

\end{multicols}

\end{document}